# Resonant TERS of a Single-Molecule Kondo System

Rodrigo Cezar de Campos Ferreira[1], Amandeep Sagwal[1,2], Jiří Doležal[1,3], Sofia Canola[1], Pablo Merino[4], Tomáš Neuman[1], Martin Švec[1,5*]

[1] Institute of Physics, Czech Academy of Sciences; Cukrovarnická 10/112, CZ16200 Praha 6, Czech Republic

[2] Faculty of Mathematics and Physics, Charles University; Ke Karlovu 3, CZ12116 Praha 2, Czech Republic

[3] Institute of Physics, École Polytechnique Fédérale de Lausanne, CH-1015 Lausanne, Switzerland

[4] Instituto de Ciencia de Materiales de Madrid; CSIC, Sor Juana Inés de la Cruz 3, E28049 Madrid, Spain

[5] Institute of Organic Chemistry and Biochemistry, Czech Academy of Sciences; Flemingovo náměstí 542/2. CZ16000 Praha 6, Czech Republic

**Abstract:**

Single-molecule tip-enhanced Raman spectroscopy (TERS) under ultra-high vacuum (UHV) and cryogenic conditions enables exploration of the relations between the adsorption geometry, electronic state, and vibrational fingerprints of individual molecules. TERS capability of reflecting spin states in open-shell molecular configurations is yet unexplored. Here we use the tip of a scanning probe microscope to lift a perylene-3,4,9,10-tetracarboxylic dianhydride (PTCDA) molecule from a metal surface to bring it into an open-shell spin one-half anionic state. We reveal a correlation between the appearance of a Kondo resonance in the differential conductance spectroscopy and concurrent characteristic changes captured by the TERS measurements. Through a detailed investigation of various adsorbed and tip-contacted PTCDA scenarios, we infer that the Raman scattering on the suspended PTCDA is resonant with a higher excited state. Theoretical simulation of the vibrational spectra enables a precise assignment of the individual TERS peaks to high-symmetry $A_g$ modes, including the fingerprints of the observed spin state. These findings highlight the potential of TERS in capturing complex interactions between charge, spin, and photophysical properties in nanoscale molecular systems, and suggest a pathway for designing spin-optical devices using organic molecules.

Precise local measurements and identification of individual vibrational states in single molecules at their natural scale of picometers are a long-standing challenge in spectroscopy. Tip-enhanced Raman spectroscopy (TERS) operated in ultra-high vacuum (UHV) has recently witnessed remarkable advances in this direction[1-3]. This technique employs a plasmonic nanocavity formed between a sharp tip apex and the substrate to enable bidirectional coupling between the electromagnetic far-field and near-field confined into subnanometric volumes, achieving extreme spatial resolution[4,5]. The TERS, along with other near-field techniques such as tip-enhanced photoluminescence (TEPL)[6-8] and electroluminescence (EL)[9-13] were integrated with the cryogenic scanning tunneling microscopy (STM) and opened the possibility to optically probe single organic molecules adsorbed on atomically flat surfaces. TERS has been applied e.g. for intramolecular mapping of vibrational modes in phthalocyanine and porphyrinoid molecules[2,3,14] and tracking conformational changes, tautomerization or deprotonation of polyaromatic hydrocarbons, using specific vibrational fingerprints[15-17]. This technique has demonstrated the ability to provide precise optical fingerprints of individual functional molecules in various chemical, electronic and charge states. Therefore application of TERS is a possible pathway to a fast and versatile tracking of the total electronic spin in the open-shell molecular configurations, which can have implications in molecule-based spin-optical device concepts. Spin states on individual molecules have been previously measured using electron transport and electron spin-resonance methods[18-24], but have not been detected optically.

Here we perform UHV-TERS experiments at an open-shell molecule, which provides an ideal playground for revealing the complex interplay among the charge, spin and electronic states of a molecule in a nanocavity and its impact on the optical fingerprints. It has been shown that a controlled atomic contact between the STM tip and molecules leads to dramatic rise of Raman intensity[25-28], which is a clear indication of enhancement beyond the electromagnetic field intensification. Therefore we select a highly-tunable single-molecule break-junction system that exhibits a reversible transition between a mixed-valence state and a singly charged radical upon decoupling from a metal substrate[29-33]. We manipulate controllably a single perylene-3,4,9,10-tetracarboxylic dianhydride (PTCDA) molecule on Ag(111) while simultaneously measuring the electronic spin state and the Raman spectra in the chemical enhancement mode. By correlating the appearance of the characteristic Kondo peak in the differential conductance spectra with the vibrational mode intensities in the chemically enhanced TERS, we identify unique Raman fingerprints of the spin state. A reference TERS mapping of a PTCDA anion on NaCl reveals the resonant character of the spectra, permitting us to perform a complete assignment of the vibrational modes to the observed spectral features based on time-dependent density functional theory (TD-DFT) calculations.

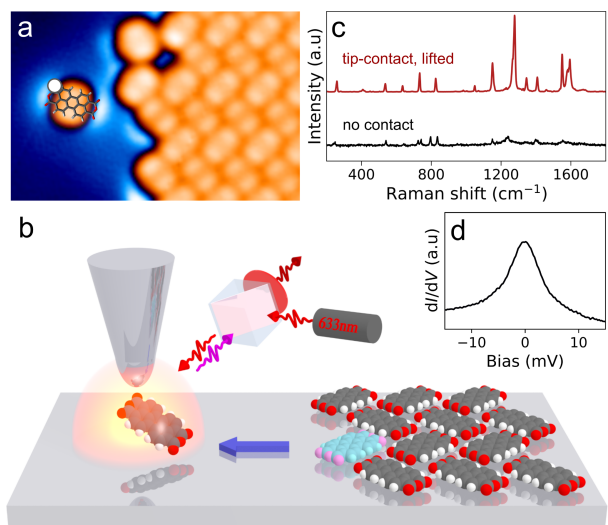

**Fig.1: Single-molecule experiment with Raman and spin measurements.** *a) Constant-current topographic image of a PTCDA molecule isolated from a 2D assembled layer by deliberate manipulation using the STM probe. b) Schematic representation of the setup used to contact and decouple the molecule from the substrate (white point in a) denotes the point of contact). Subsequently, the enhancement effect of the nanocavity is applied to detect c) the Raman fingerprint of a single-molecule contacted with the tip (red, 3 sec, 1 mV, 1 nA transport current). A spectrum for a molecule on Ag without the tip contact is provided for comparison (black, 50 sec, 1 mV, 200 pA tunneling current). d) dI/dV spectroscopy around the Fermi level on the contacted and lifted molecule, showing a peak characteristic of the spin 1/2 state.*

The setup and methodology of the experiments are schematically shown in Figs.1 and S4. PTCDA molecules are evaporated thermally onto the Ag(111) surface, using the procedure described elsewhere[31]. On-surface diffusion of the adsorbates promotes the formation of self-assembled islands of molecules. An optically active and atomically-sharp Ag-tip is prepared by controlled indentations and voltage pulses, in order to obtain a strong plasmonic response of high intensity spectrally matching the energy of the excitation source (632.8 nm HeNe laser), see Fig.S5, and the region of the Raman shift.[34] Through manipulations we extract a single PTCDA molecule from the edge of an island and transfer it onto a clean metal area. To this end, we contact the molecule *via* one of the four carbonyl terminations and drag it laterally away from the island. Subsequently the molecule is released by retracting the tip several nm and inspected by taking an STM image of the area (Fig.1a). After recontacting the molecule we detach it gradually from the surface by a controlled vertical displacement of the tip, while simultaneously measuring Raman spectra using a confocal arrangement (Fig.1b,c). For every step, a differential conductance (dI/dV) curve is recorded, which is indicative of the electronic spin state of the molecule by showing a prominent Kondo peak centered at the Fermi level when the molecule has a single unpaired electron (Fig.1d)[31]. The molecule bonded to the tip apex can be typically lifted and lowered repeatedly for several cycles in a reproducible manner, as shown in Figs. S6 and S7. The Raman signal of a lifted PTCDA is about two orders of magnitude more intense compared to the Raman signal obtained in the tunneling regime over the molecule on the surface (Fig.1c).

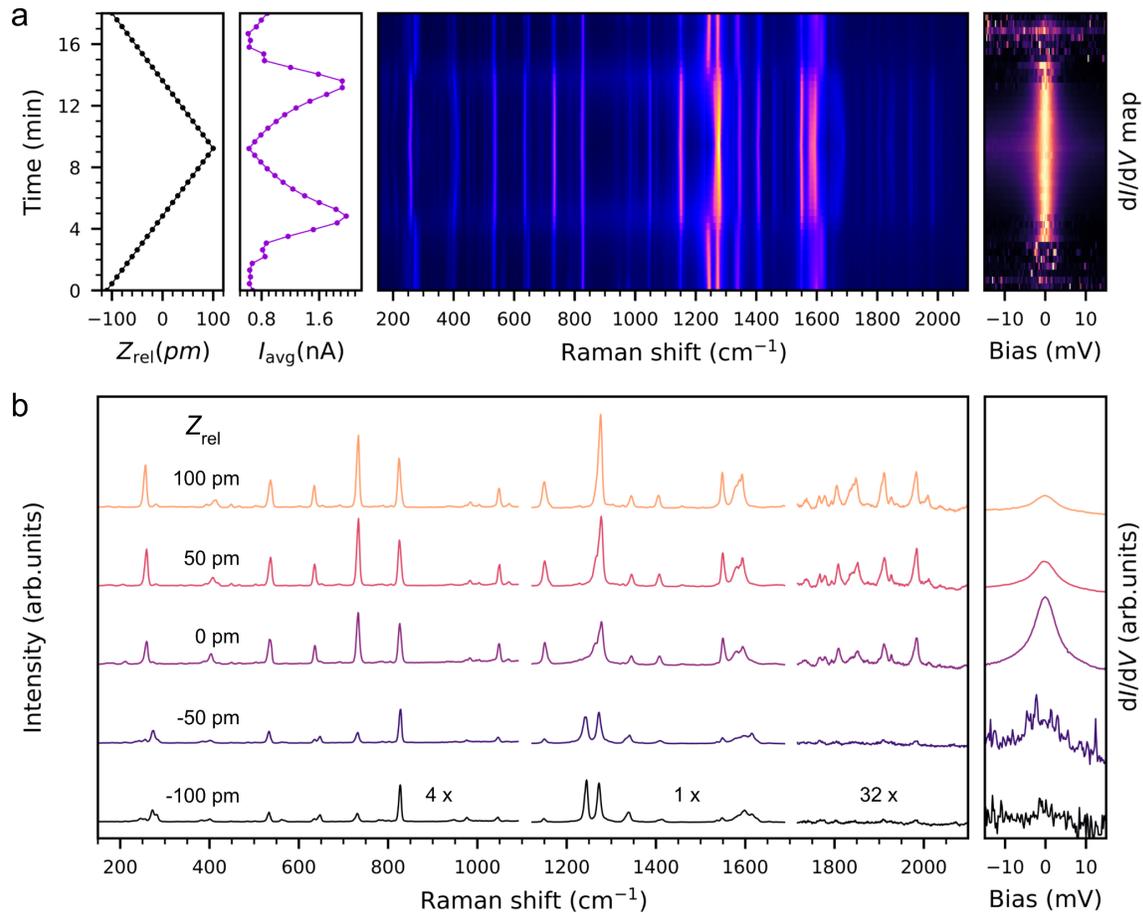

**Fig.2: Tracking the spin transition with Raman.** *a) Current, TERS and dI/dV intensity (normalized) as a function of the PTCDA height of lifting ($Z_{rel}$) from the Ag(111) substrate, relative to the onset of the Kondo signature. The lifting step size was 10 pm. b) detailed Raman and dI/dV spectra (non-normalized) at selected heights around the spin transition.*

Fig.2a shows a complete cycle of the lifting-laying experiment, with the measured current, TERS and d$I$/d$V$ as a function of the relative height ($Z_{rel}$). The as-contacted PTCDA on Ag(111) is still in a mixed valence state, and according to previous analyses its LUMO is situated below the Fermi level of the substrate and nearly doubly occupied[29,36]. An effective removal of this mixed valence at the lifting height ($Z_{rel} = 0$) is hallmarked by a sharp increase of the conductivity around zero bias voltage. The molecule becomes a single-electron radical (*S = 1/2*), due to the upshift of the LUMO through the Fermi level[31]. Varying the $Z_{rel}$ around this setpoint from -110 to 100 pm and back (see Fig.2a), we trace the evolution of all the signals simultaneously. The TERS shows reversible transformations coincident with the conductivity maxima, in particular remarkable changes of the 1243 cm$^{-1}$ line intensity (see Fig.1b), that strongly correlate with the appearance and disappearance of a strong, clean Kondo signature in the normalized d$I$/d$V$ spectra. Such behavior points out a relation between the changes in the TERS spectrum and the electronic state before and after the spin transition. In addition, with further increasing $Z_{rel}$ a decay of the conductivity occurs due to a progressive decoupling of the PTCDA from the substrate, while the TERS overall intensity increases at the same time due to the progressing alignment of the transition dipole with the nanocavity polarization.

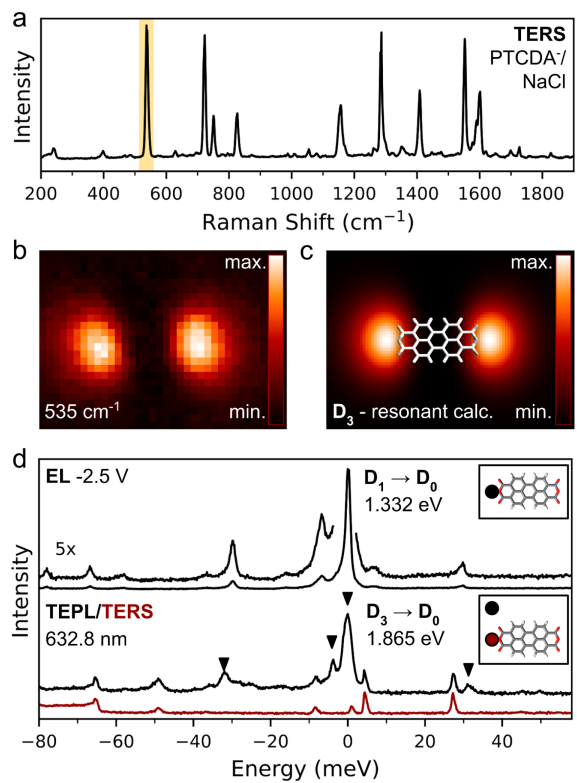

**Fig.3: Submolecular mapping of Raman intensity under resonant conditions.** *a) Raman spectrum of a single PTCDA$^-$ on NaCl, acquired at 0.5 V above the oxygen-termination of the PTCDA. b) Raman intensity map taken from the region around 535 cm$^{-1}$ (denoted by yellow band in a) in the constant height mode and normalized. Size 3.6 x 2.5 nm$^2$, 0.5 V. c) Simulated Raman map, resonant with $D_3$-state. d) Comparison of the EL spectrum[10] corresponding to the $D_1 \rightarrow D_0$ transition and TEPL spectrum detected using the 632.8 nm excitation, corresponding to the $D_3 \rightarrow D_0$ transition. The background Raman spectrum is plotted in red color below on the same energy scale for reference. The dots in the insets show the approximate locations of the measurements. Black arrowheads denote the peaks attributed to the TEPL contribution. Both TEPL and Raman spectra have the corresponding far-field contribution subtracted.*

In order to elucidate the regime of the observed intense TERS spectra, to provide grounds for a theoretical simulation, and to support the link of the spectral fingerprints to particular states of the molecule, we performed a reference measurement on the flatly adsorbed PTCDA, decoupled from the Ag substrate by two NaCl layers and without the tip contact (Fig.3a). In this configuration, the molecule also adopts a singly-charged anionic state (PTCDA⁻), as confirmed by previous works[10,37]. Fig.3d shows that a relatively strong TERS signal is present, similar to the Raman spectrum of the suspended molecule in the *S = 1/2* state (for comparison see Fig.4a). Although the amplitudes of the two configurations differ, especially in the range below 1000 cm$^{-1}$, likely due to their specific geometries and local environments, they share a common feature - the diminished peak at 1243 cm$^{-1}$, in contrast to the scenarios where the PTCDA is coupled to the metal substrate and does not have an open-shell character.

The Raman signal is adequately strong to take a spatially-resolved TERS of the single anion. The intensity map for the prominent line at 535 cm$^{-1}$ (marked in yellow) in Fig. 3b, corresponding to the transversal stretching mode (see below), shows a pattern consisting of two lobes localized at both ends of the molecule with the carboxylic terminations. Maps in Fig.S8 at the energies of all other dominant peaks manifest a nearly identical characteristic pattern, which resembles the electroluminescence distribution, previously reported for the PTCDA⁻ first excited state transition to the ground state ($D_1 \rightarrow D_0$)[10]. This suggests a resonant character of the Raman scattering. When resonantly excited, the Raman intensity is proportional to the Franck-Condon activities of the corresponding vibrations, and its tip-position-dependence is modulated proportionally to the coupling of the nanocavity to the resonant excited state[38,39]. However since the energy of the $D_1 \rightarrow D_0$ transition is 1.33 eV, it cannot contribute resonantly to the Raman scattering at the excitation wavelength 632.8 nm (1.96 eV). The energy of the transition from the second excited state to the ground state ($D_2 \rightarrow D_0$), being 1.54 eV, is also detuned from the excitation and its transition dipole moment is oriented along the molecular short axis, perpendicular with respect to the observed TERS pattern[10]. Therefore we expect a higher excited state with an energy close to the incident light to be responsible for the observed resonant character of the TERS.

In the off-axis position in the peripheral region of the molecule we are able to detect a prominent spectral feature around 1.865 eV, broader than the rest of the Raman peaks, accompanied by longitudinal stretching mode sidebands at ±31.5 meV, also found in the high-resolution $D_1 \rightarrow D_0$ electroluminescence spectrum (plotted for comparison in Fig.3d). Based on our TD-DFT calculations, which place the third excited state ($D_3$) approximately 290 meV above the $D_2 \rightarrow D_0$ transition, we attribute the 1.865 eV peak and its sidebands to the $D_3 \rightarrow D_0$ transition of the PTCDA⁻. This transition has the dipolar moment oriented along the longitudinal axis of the molecule, in accord with the TERS emission pattern in Fig.3b and a remarkably high oscillator strength, stemming from a constructive interference between the dominant electronic transitions involved in the excitation to the many-body $D_3$ state, as explained in detail in the Supplementary Information[35]. A resonant simulation involving the $D_3 \rightarrow D_0$ transition density and the point-charge approximation of the nanocavity agrees exceptionally well with the experimental TERS map (shown in Fig.3c). The corresponding simulated $D_3$-resonant Raman spectrum of a free-standing PTCDA⁻ (see Fig.4c) is dominated by the highest-symmetry *$A_g$* modes, in accord with the previously made assessment[40,41]. The relative Raman activities of the vibrational modes give a convincing agreement with the experimental TERS of the suspended PTCDA⁻ in the *S = 1/2* state, allowing an assignment of the individual vibrational modes and extrapolation to the other measured configurations.

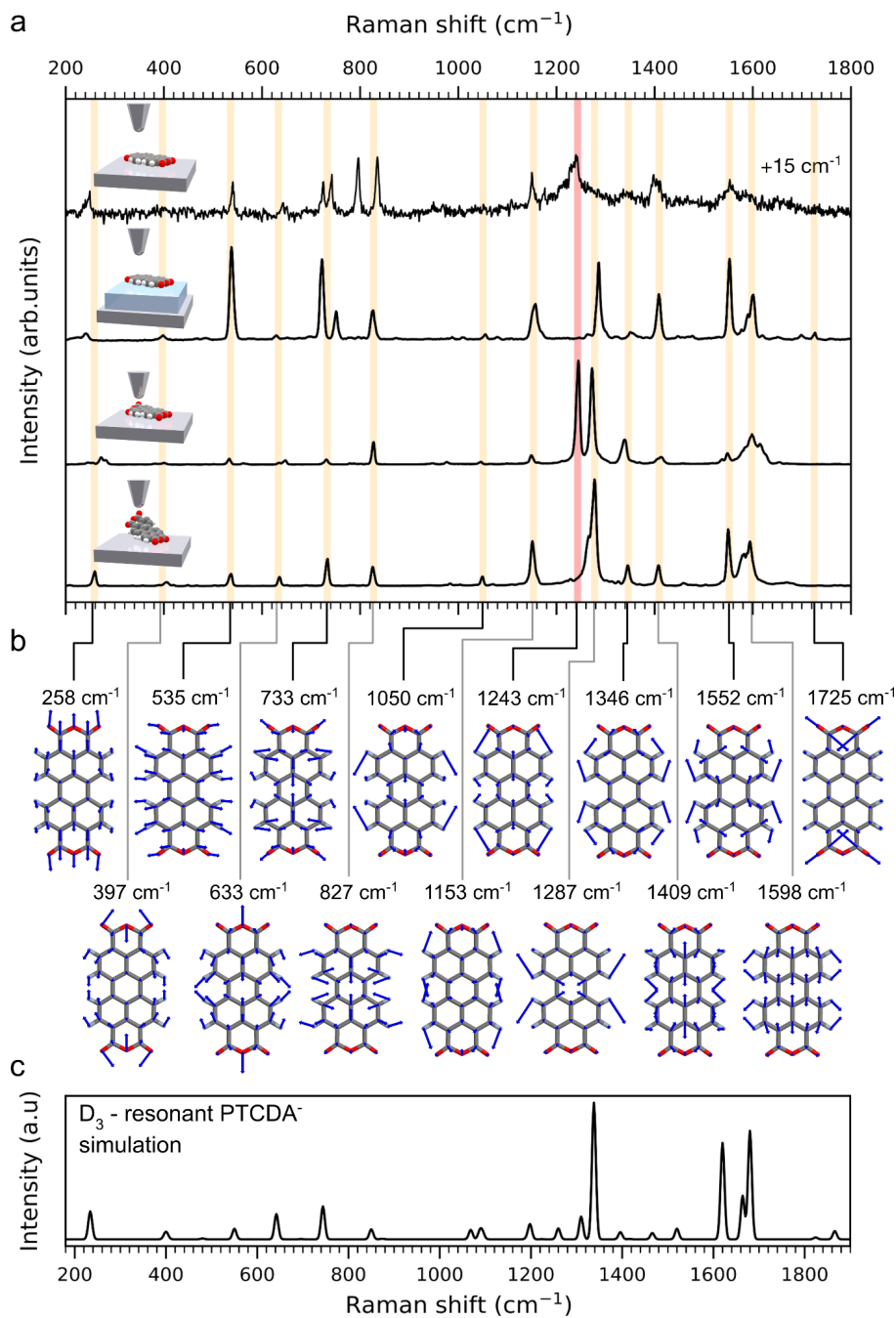

**Fig.4: Comparison of observed peaks and assignment to vibrational modes.** *a) Top-down comparison of the spectra taken on PTCDA when adsorbed on Ag, on NaCl/Ag, on Ag upon contact with the probe and when lifted. The spectrum of the PTCDA adsorbed on Ag is shifted by 15 cm$^{-1}$ b) The dominant $A_g$ modes, assigned to the spectral features, based on the agreement between the experimental spectrum in a) and theoretical simulation in c). c) The theoretical Raman spectrum of a PTCDA⁻ single radical calculated in resonance with the $D_3$ state.*

Low-wavenumber modes in the spectra at 258 $cm^{-1}$ and 535 $cm^{-1}$ in Fig.4 correspond to the stretching of the molecule in the longitudinal and transversal directions, respectively. The shift in the longitudinal mode frequency during lifting reflects the relaxation of the PTCDA backbone occurring concurrently with the spin transition and the development of anionic state (detail visible in Fig.2b). The longitudinal mode is also softened in the Ag- and NaCl-adsorbed configurations, probably due to a bond expansion along the corresponding coordinate, stemming from interaction with the substrates. For the molecule on NaCl, the Raman activities of the transversal stretching at 536 $cm^{-1}$ and breathing at 733 $cm^{-1}$ are strongly increased. The latter is also softened to 715 $cm^{-1}$ in both of the adsorbed scenarios, and accompanied by an emerging lowered-symmetry mode 33 $cm^{-1}$ higher in energy. As we show in the Supplementary Information[35], the rise in activities of the high-symmetry modes can be understood in terms of the Franck-Condon picture. We can therefore infer that upon transition to $D_3$, in comparison to the other configurations, the molecule on NaCl is undergoing a larger distortion along these particular modes.

The most prominent spectral lines at 1243, 1287 and 1552 $cm^{-1}$, are all vibrational modes with a significant C-H bending character. The line at 1243 $cm^{-1}$ (denoted with a red bar in Fig.4a) is the only visible peak that is significantly reduced (and shifted to higher frequency) during the development of the single-anionic state and can be considered as a hallmark of the PTCDA spin transition. Conversely, the 1287 $cm^{-1}$ is present in all situations where the spectra are chemically enhanced due to contact with the tip or resonant, *i.e.* always except for the molecule on Ag. Other less intense vibrational peaks are common in all the probed cases at 1153 and 1346 $cm^{-1}$ and can be attributed to the C-H scissoring and bending modes. Higher-order longitudinal stretching peaks appear at 1409 $cm^{-1}$ and 1598 $cm^{-1}$. Interestingly, the C-O stretching found at 1725 $cm^{-1}$ for the unperturbed PTCDA⁻ on NaCl, is shifted about 60 $cm^{-1}$ lower and significantly broadened for the lifted molecule. This is an indication of the C-O bond weakening due to the chemical interaction of the oxygen atoms with the tip and the metal substrate.

In summary, we have studied the TERS spectrum of single PTCDA molecule in four basic configurations of various electronic and adsorption states: (i) adsorbed on metal in a mixed valence state, (ii) adsorbed on a decoupling NaCl bilayer in open-shell *S = 1/2* state (iii) adsorbed on metal and contacted by the tip still in the mixed valence state and (iv) suspended between the Ag(111) and the tip with the *S = 1/2*. We have found that the TERS signals of the PTCDA measured with a conventional 632.8 nm laser excitation are significantly increased by physical contact with the tip and by the Raman scattering resonant with the $D_0 \leftrightarrow D_3$, which was found by TEPL. Comparison of the experimental TERS of the *S = 1/2* state of the singly charged PTCDA anion with the resonant-Raman TD-DFT calculations yields a good agreement allowing an unambiguous identification of the Raman-active modes. By following the evolution of the Raman activities of individual modes in the spectra during lifting/laying of the molecule from/to the metal substrate and comparing to a PTCDA⁻ adsorbed on NaCl, we have concluded that the appearance of Kondo resonance (a clear hallmark of the *S = 1/2* state) anticorrelates with the high intensity of a C-H bending mode at 1243 cm$^{-1}$. Therefore this particular mode can be used as an indicator of the spin state. With further lifting of the PTCDA$^-$ from the metal substrate, the overall TERS intensity is increasing due to the progressing alignment of the transition dipole with the nanocavity polarization. The results of this work bring the prospects of measurements when a higher excited state resonance, tip enhancement or tip-contact chemical enhancement effect will work in combination and result in genuinely high Raman yields on single molecules, while preserving the intrinsic Raman activities of individual vibrational modes. We envisage the exciting possibility that characteristic changes in single-molecule Raman spectra as reported here could be engineered and employed in organic-based spin-optical transducers.

**Methods:**

The experiments were performed in a LT-STM (Createc Gmbh) operating at 7K and in an ultrahigh vacuum (UHV) environment below 5×10$^{-11}$ mbar base pressure. The Ag(111) single crystal was prepared by standard cycles of Ar$^+$ sputter and annealing at 550°C. To obtain islands with 2-4 layers of NaCl, it was thermally evaporated at 610°C on the surface held at 120°C during 3-4 min. PTCDA molecules were sublimated at 350°C onto both Ag(111) and NaCl/Ag(111) samples, kept at room temperature or at 5K, respectively. The optical setup was based on a confocal arrangement. d$I$/d$V$ curves were measured with lock-in technique using modulation amplitude 2 mV. Further details of the experimental methods are described in the Supplementary Information[35].

We modeled the spectral response and photon map using time-dependent density functional theory (TDDFT) as implemented in Gaussian 16[42]. The spectral response was modeled using the resonance Raman approach considering both the Herzberg-Teller and Franck-Condon activities of the vibrations[43] (further details are in the Supplementary Information[35], where we discuss the resonance Raman spectra calculated for the $S_0 \leftrightarrow S_1$, $D_0 \leftrightarrow D_1$ and $D_0 \leftrightarrow D_3$ transitions). The photon map of vibronic peaks was calculated assuming that a single plasmon mode of the tip couples to the molecular $D_0 \leftrightarrow D_3$ transition. Details of the implementation are also discussed in the Supplementary Information[35].

**Acknowledgements**: RCCF, MS, AS and JD acknowledge the funding from the Czech Science Foundation grant no. 22-18718S and the support from the CzechNanoLab Research Infrastructure supported by MEYS CR (LM2023051). TN and SC acknowledge the Lumina Quaeruntur fellowship of the Czech Academy of Sciences. Computational resources were supplied by the project "e-Infrastruktura CZ" (e-INFRA CZ LM2018140) supported by the Ministry of Education, Youth and Sports of the Czech Republic. PM acknowledges financial support from projects RYC2020-029800-I, EUR2021-122006, PID2021-125309OA-I00, CNS2022-135658, and TED2021-129416A-I00 funded by MCIN/AEI/10.13039/501100011033.

# Supplementary Information for

# Resonant TERS of a Single-Molecule Kondo System

Rodrigo Cezar de Campos Ferreira[1], Amandeep Sagwal[1,2], Jiří Doležal[1,3], Sofia Canola[1], Pablo Merino[4], Tomáš Neuman[1], Martin Švec[1,5*]

[1] Institute of Physics, Czech Academy of Sciences; Cukrovarnická 10/112, CZ16200 Praha 6, Czech Republic

[2] Faculty of Mathematics and Physics, Charles University; Ke Karlovu 3, CZ12116 Praha 2, Czech Republic

[3] Institute of Physics, École Polytechnique Fédérale de Lausanne, CH-1015 Lausanne, Switzerland

[4] Instituto de Ciencia de Materiales de Madrid; CSIC, Sor Juana Inés de la Cruz 3, E28049 Madrid, Spain

[5] Institute of Organic Chemistry and Biochemistry, Czech Academy of Sciences; Flemingovo náměstí 542/2. CZ16000 Praha 6, Czech Republic

## Theoretical calculations

### Time-dependent density-functional-theory calculations

To theoretically describe the resonance Raman spectra we perform time-dependent density-functional theory (TDDFT) modeling of the optical response of the PTCDA molecule. Our approach neither does take into account geometrical deformations of the molecule due to its attachment to the tip and the substrate nor it considers the electronic coupling of the molecule with the metallic electrodes, as the description of such phenomena is beyond the scope of the present work. We therefore resort to calculation of the molecule optical response in a vacuum, considering that upon lifting, the electronic structure of the molecule corresponds to the one of its fully decoupled counterpart. In particular we perform TDDFT calculations using Gaussian 16 revision A.03 and C.01[1] using the double zeta Gaussian basis set including polarization functions 6-31G* and the range-separated hybrid functional $\omega$B97XD[2].

We first optimize the molecular geometry in the ground state and calculate the molecular vibrations considering that the ground state is (i) a neutral singlet ($S_0$) and (ii) a negative doublet ($D_0$). We perform a geometry optimization and vibrational analysis in selected excited states of the molecule. For excited state $S_1$ of the neutral molecule and states $D_1$, $D_3$ of the negative molecule, we calculate resonance Raman response using the Franck-Condon and Herzberg-Teller (FCHT) analysis implemented in Gaussian 16. In particular, since the molecule has been shown to be singly negative when decoupled by lifting from a metal substrate[3,4], we compare the experimentally observed spectra to the electronic transitions between the ground state and the first and third excited state ($D_1^-$ and $D_3^-$) which are the two lowest lying states featuring a transition dipole moment aligned along the PTCDA long axis. We also calculate the Raman spectrum in the first excited ($S_1$) state of the neutral molecule.

We summarize the results of the calculations of the electronic structure in Fig.S1. Figure S1a shows a schematic diagram of the many-body ground and the first excited singlet state of the neutral molecule. The energy of the transition evaluated in the ground-state geometry is shown in the diagram (or in the excited state geometry in the brackets). The corresponding calculated transition density of the $S_0 \leftrightarrow S_1$ transition is shown. The transition density shows a dipolar moment along the molecule's long axis. Alongside with the many-body states we show their corresponding orbital occupations in the configuration diagram in the inset. The ground-state configuration and orbitals are derived from the underlying DFT calculation. For the excited state, we derive the configuration from the dominant electron-hole pair transitions contributing to the excitation as calculated by the linear-response TDDFT. The orbitals (highest-occupied - HOMO, lowest unoccupied - LUMO, LUMO+1 and LUMO+2) occupations corresponding to the $S_0$ state are shown on the right for completeness. From the diagrams we see that the HOMO-LUMO transition dominantly contributes to $S_0 \leftrightarrow S_1$.

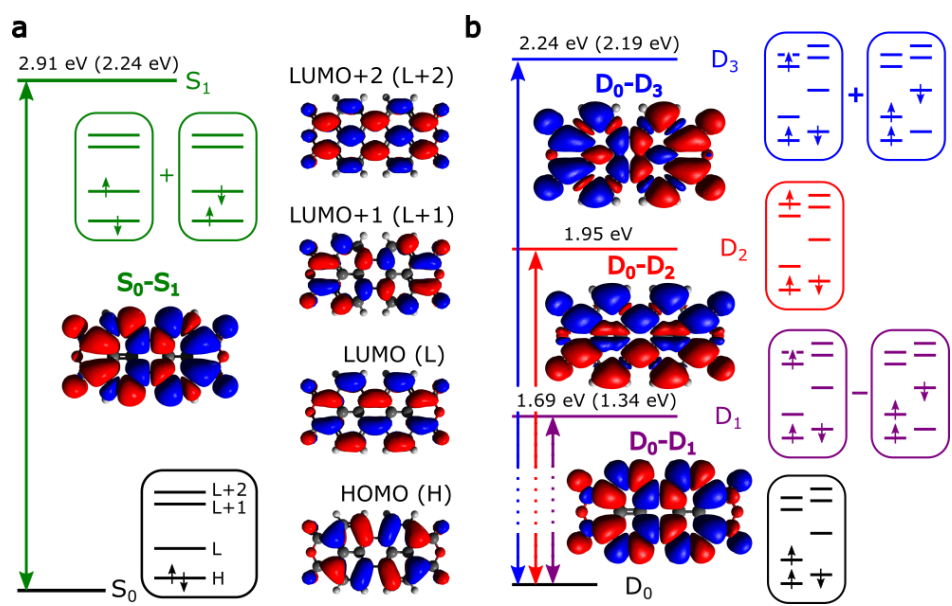

Fig.S1: *Electronic structure of the (a) neutral and (b) negative PTCDA molecule. (a) The transition density of the $S_0 \leftrightarrow S_1$ transition is shown alongside the diagram showing the excitation energy and the schematic electron configuration diagram of the electronic states. The molecular orbitals (HOMO - H, LUMO - L, LUMO+1 - L+1, and LUMO+2 - L+2) obtained for the neutral molecule are shown next to the many-body diagram. (b) Excitations of the negative molecule with the corresponding transition densities, excitation energies, and electron configuration diagrams. The energy labels in the diagram mark the energies calculated as vertical transitions in the ground-state (and in the excited-state in the brackets) geometry.*

Figure S1b displays the electronic structure of the negative molecule. The first three excited states are shown with the corresponding transition energies in the ground-state (excited-state) geometry labeled in the diagram. The three corresponding transition densities ($D_0 \leftrightarrow D_1$ - purple label, $D_0 \leftrightarrow D_2$ - red label, $D_0 \leftrightarrow D_3$ - blue label) are shown. The transition densities corresponding to $D_0 \leftrightarrow D_1$ and $D_0 \leftrightarrow D_3$ feature a dipole moment oriented along the long axis of the molecule, in contrast to $D_0 \leftrightarrow D_2$ that has a dipole moment along the short axis of the molecule. We also show the configuration diagram showing the orbital occupations in the doublet states. In the open-shell configuration we distinguish the orbitals for the spin up and spin down electrons as the underlying DFT calculation uses the spin-unrestricted ansatz and thus the orbitals are calculated independently for the two spin channels. Despite this, the orbitals can still be associated with the ones of the neutral molecule shown in Fig.S1a. The excited-state configurations are again derived from the dominant electron-hole pair transitions contributing to the respective excitations. The configuration diagrams are color-coded using the same labeling scheme as for the transition densities. From the configurations it is apparent that the $D_0 \leftrightarrow D_1$ and $D_0 \leftrightarrow D_3$ transitions contain the same electron-hole pairs, but the electron-hole pair superposition differs in sign. As a consequence, the transition dipole moment of the $D_0 \leftrightarrow D_3$ is significantly larger than that of the $D_0 \leftrightarrow D_1$ transition. This is because in the first case the contributing electron-hole pairs superpose constructively, whereas in the latter case destructively. The $D_0 \leftrightarrow D_3$ transition can thus be expected to significantly contribute to the optical response of the molecule, including the (near) resonance-Raman response at the energy of the laser used in the experiments.

We now focus on the calculation of resonance Raman spectra corresponding to laser tuned close to the $D_0 \leftrightarrow D_3$ transition. Since in the experiment the incident laser may not be exactly resonant with the electronic transition, we test the effect of the incident-laser detuning and calculate the resonance Raman spectra for several frequencies of the incident laser. The results normalized to the maximum are shown in Fig.S2a in the form of a waterfall plot. The spectral peaks vary their relative intensities as the laser frequency is tuned and we observe that particularly the peaks of frequency <1000 cm$^{-1}$ have relatively larger intensity in the range of incident laser frequencies between approximately 16000-19000 cm$^{-1}$ and their relative weight in the spectrum peaks around 17800 cm$^{-1}$. We also see that the peaks between 1600 cm$^{-1}$ and 1700 cm$^{-1}$ vary their relative intensity when the laser frequency is detuned. We note that the spectrum at 17000 cm$^{-1}$ is used in the main text as it is the best match with the experiment.

In Fig.S2b-d we compare the resonance Raman spectra calculated for the three electronic transitions: (b) $D_0 \leftrightarrow D_1$, (c) $D_0 \leftrightarrow D_3$, and (d) $S_0 \leftrightarrow S_1$. We observe that the spectra for the respective transitions differ considerably. The Raman peaks of the lower-energy vibrations are more pronounced in the $D_0 \leftrightarrow D_3$ spectrum than in the $D_0 \leftrightarrow D_1$ and spectra $S_0 \leftrightarrow S_1$. This finding supports the interpretation of the experimental spectrum as a resonance Raman spectrum of the $D_0 \leftrightarrow D_3$ transition.

In Fig.S3 we show the resonance Raman spectra of the three transitions: (a) $D_0 \leftrightarrow D_1$, (b) $D_0 \leftrightarrow D_3$, and (c) $S_0 \leftrightarrow S_1$. They have been calculated using the full Franck-Condon-Herzberg-Teller formalism (black lines), and the spectra calculated only using the Franck-Condon activity of the vibrations (red lines). We see that up to smaller spectral contributions, the spectra are well described as Franck-Condon spectra of the respective transitions.

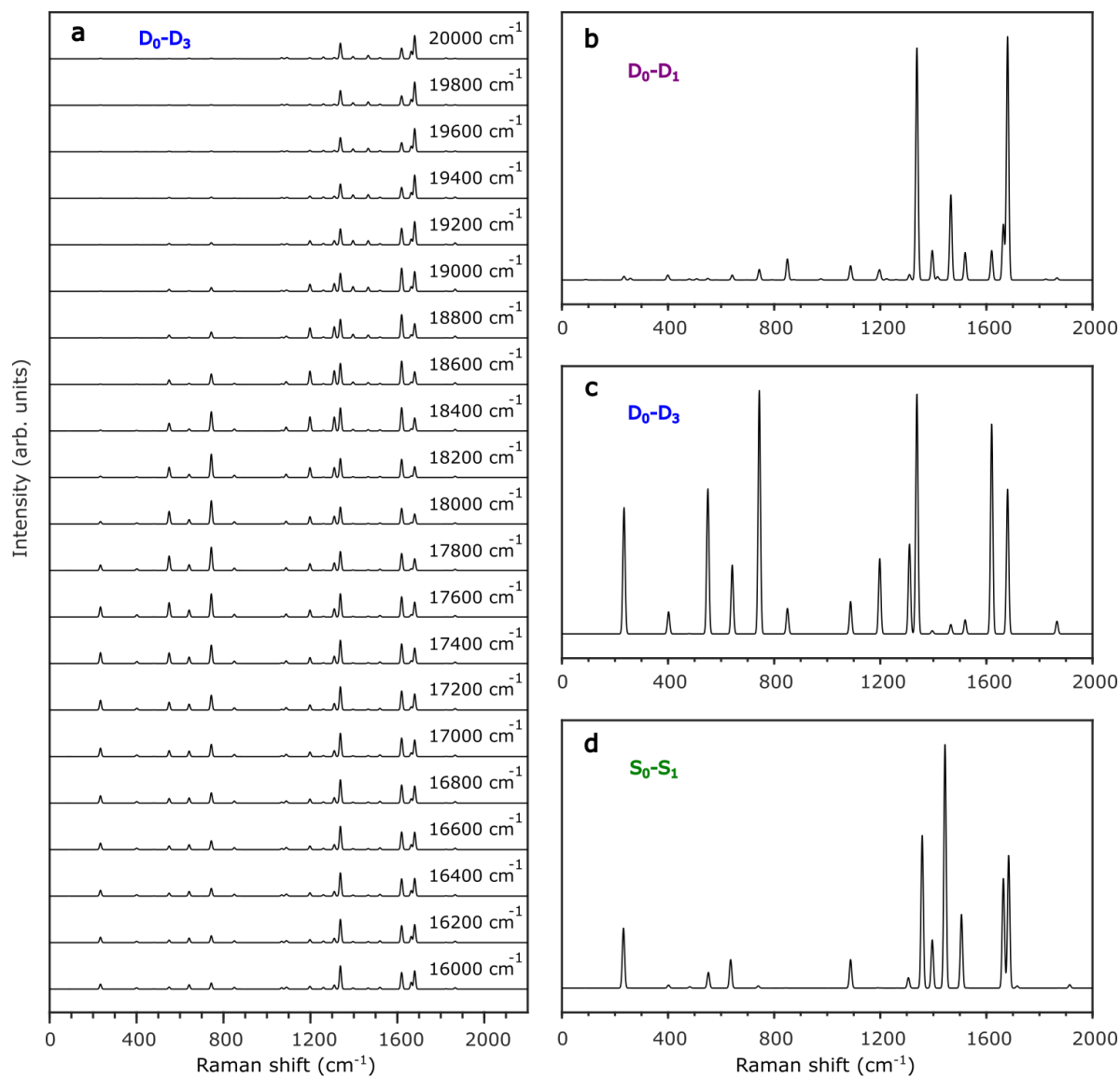

Fig.S2: *Near-resonance Raman spectra of PTCDA corresponding to different electronic transitions. (a) Spectra corresponding to Raman spectra of the $D_0 \leftrightarrow D_3$ transition assuming different excitation frequencies (the 0-0 transition being at 17452.56 cm$^{-1}$). (b)-(d) Resonance Raman spectra assuming exact tuning of the excitation laser with the (b) $D_0 \leftrightarrow D_1$ transition, (c) $D_0 \leftrightarrow D_3$ transition, and (d) $S_0 \leftrightarrow S_1$ transition.*

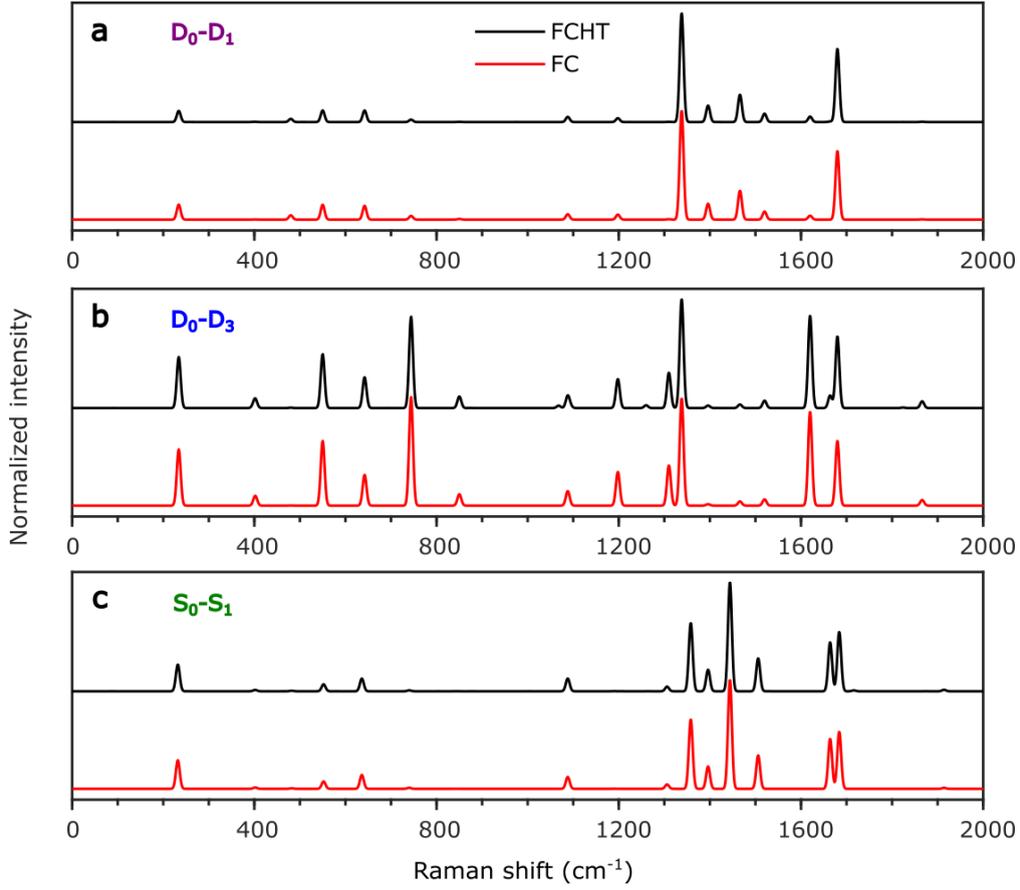

Fig.S3: *Comparison of Raman spectra calculated using the full Franck-Condon-Herzberg-Teller approach (FCHT - black lines) and spectra calculated using only the Franck-Condon principle (FC - red lines). The spectra were calculated assuming the exact tuning of the laser to the transition energy of the (a) $D_0 \leftrightarrow D_1$ transition, (b) $D_0 \leftrightarrow D_3$ transition, and (c) $S_0 \leftrightarrow S_1$ transition.*

**Calculation of the Raman map**

In Fig.3c of the main text we show the calculated Raman photon intensity map (Raman map) recorded as a function of the lateral position of the tip. We calculate the Raman map assuming that the $D_0 \to D_3$ transition is close to the resonance with the excitation energy. In this situation the Raman signal recorded as a function of the position of the tip $r_{tip}$ can be approximated as $I_R(r_{tip}) \propto |g_{pl}(\omega_L, r_{tip})|^2 |g_{pl}(\omega_{St}, r_{tip})|^2 I_0$, where $I_0$ is the intensity of the incident laser, and $g_{pl}(\omega, r_{tip})$ is the plasmon-exciton coupling that we calculate as $g_{pl}(\omega, r_{tip}) \propto \int \phi_{pl}(r - r_{tip}; \omega) \rho_{D_0 \to D_3}(r) d^3r$, with $\phi_{pl}(r - r_{tip})$ being an approximation of the quasi-static electric potential of the plasmon mode (representing its mode profile), and $\rho_{D_0 \to D_3}(r)$ being the transition density of the $D_0 \to D_3$ transition. We would like to point out that in the expression for $I_R(r_{tip})$, one commonly finds the plasmon-enhancement factors $f_{pl}$ (defined as $f_{pl} = |E_{loc}|/|E_{inc}|$ with $E_{inc}$ and $E_{loc}$ being the incident and local electric field, respectively) instead of the plasmon-exciton coupling. The use of the factor $f_{pl}$ is justified when the molecular response can be treated in the dipolar approximation. In the case of plasmon-exciton coupling in a STM and considering near-resonant Raman[5,6], it is the coupling of the transition density of a particular excitonic transition with a particular localized plasmonic mode treated beyond the dipole approximation that determines the scattering properties of the molecule (assuming that the Franck-Condon mechanism is at play). The plasmon-exciton coupling should also be generally evaluated at the frequency of the incident light $\omega_L$ and at the frequency of the Stokes photon $\omega_{St}$. However, with the assumption that the plasmonic mode profile does not significantly vary with frequency (i.e. that a single plasmon mode is being excited and the same mode participates in the emission process) we obtain the simplified expression $I_R(r_{tip}) \propto |g_{pl}(r_{tip})|^4$ that we use to calculate the map in Fig.3c. We approximate the plasmonic mode profile by the potential of a pair of point charges of equal magnitude and opposite sign positioned 1.5 nm above and 2.5 nm below the plane of the molecule, respectively, at the same lateral position. This potential sufficiently approximates the electric field of a localized gap plasmon in the gap; we note that the details of the field distribution (beyond the field localization) do not play a significant role in the resulting shape of the photon map.

**Experimental details**

**The optical setup and nanocavity plasmon tuning**

In the optical setup used for the experiments, schematically depicted in Fig.S4, the excitation source was a He-Ne (632.8 nm) continuous-wave laser, collimated with an *f* = 15 mm lens, guided through a ND filter, half-wave plate, polarizer and a noise eater, to form a stable-intensity beam with polarization along the tip-sample axis. Typical total beam power used for the experiments varied from 50-200 uW. The beam was focused into the SPM junction by another internal SPM lens (also with *f* = 15 mm). The outgoing light from the SPM nanocavity was filtered using a 633 nm bandpass edge filter. The TERS and TEPL spectra were measured in cumulative mode by an Andor Kymera 328i spectrograph with a 1200 grooves/mm, 500 nm blaze grating, connected to a custom control computer and homemade control software based on the Andor SDK. The Ag tips made of 25 μm diameter Ag wire were sharpened by focused $Xe^+$ ion beam. Further cleaning by head-on $Ar^+$ sputtering was done before insertion into the SPM head and final shaping was done using nanoindentations and voltage pulses.

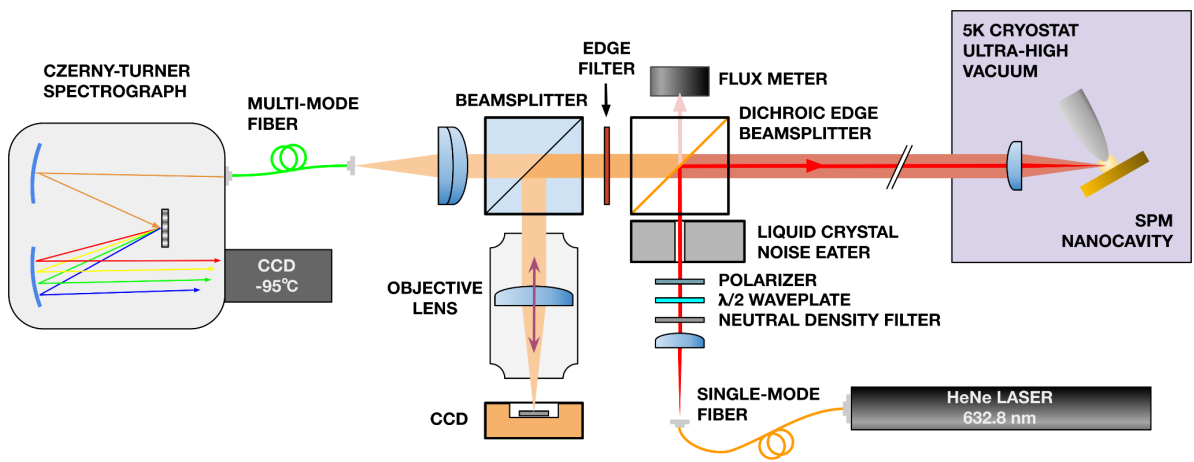

Fig.S4: *The scheme of the confocal optical setup for the measurements of single-molecule TERS, TEPL and EL at 5K and cryogenic conditions.*

The nanocavity frequency spectrum was determined by electroluminescence and was tuned to cover the range of the excitation energy and the Raman scattering. Test of the photon coupling to the nanocavity was performed using the field emission resonance measurements with a lock-in technique. As the efficient coupling causes electron energy conversion, it manifests itself as a downshift of the electron tunneling resonances, corresponding to the excitation source energy (see Fig.S5)[7].

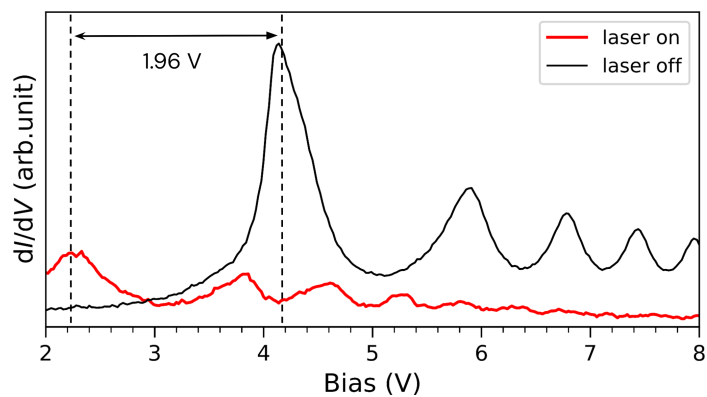

Fig.S5: *Field-emission resonances, measured as dI/dV at constant current feedback with an Ag tip on a clear Ag(111) surface as a function of the irradiation being on/off. The current setpoint was 100 pA, lock-in modulation amplitude 50 mV.*

**Reproducibility of the lifting data**

We performed several sessions of the lifting experiments, with various tips and PTCDA molecules. We have typically achieved reproducibility of the measured dI/dV and Raman spectra over more than two complete cycles. Figs. S6 and S7 show additional two datasets, measured with the 600/nm grating (at lower resolution with respect to the data presented in Fig.2 which was measured with the 1200/nm grating for high-resolution).

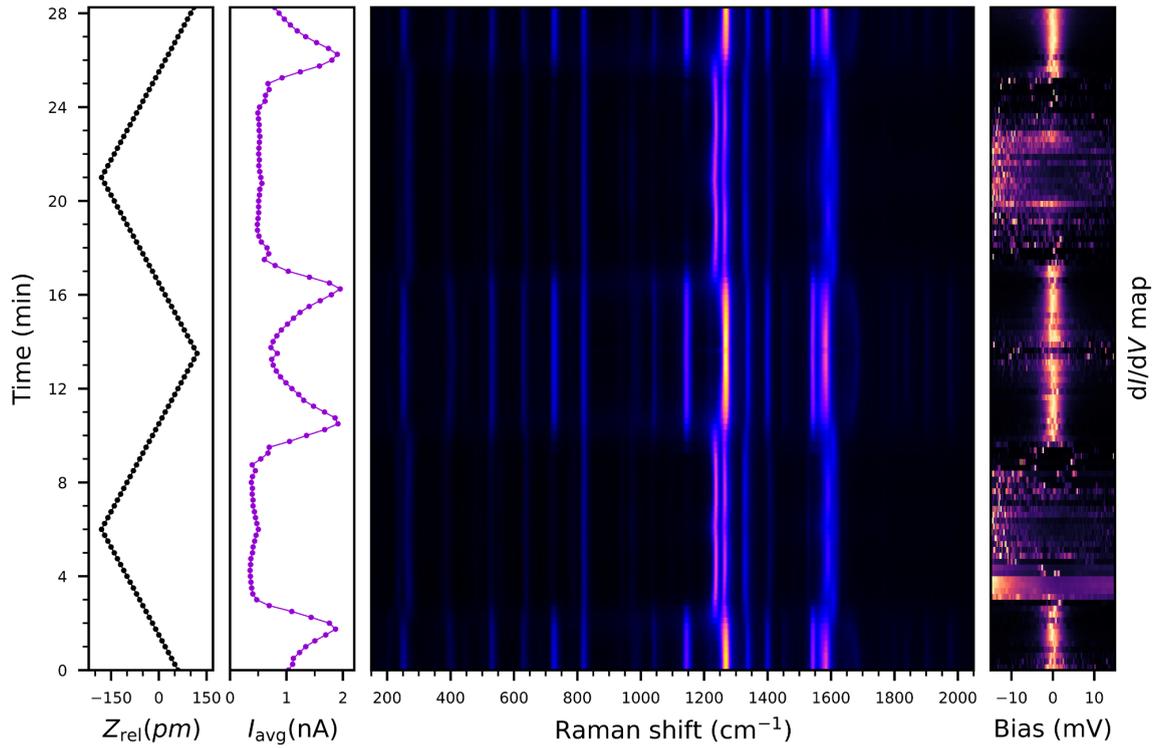

Fig.S6: *Current ($I_{avg}$), TERS and normalized dI/dV intensity (from left to right) as a function of the PTCDA height of lifting ($Z_{rel}$) from the Ag substrate, relative to the onset of the Kondo signature. The lifting step size was 10 pm.*

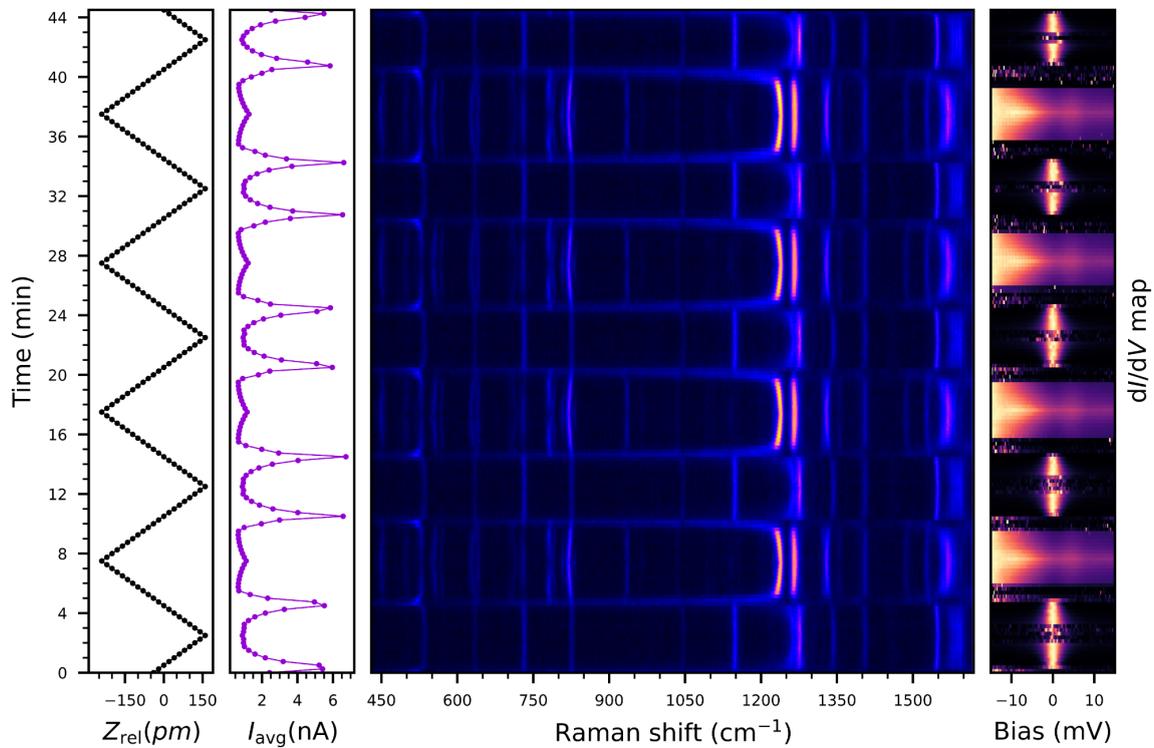

Fig.S7: *Current ($I_{avg}$), TERS and normalized dI/dV intensity (from left to right) as a function of the PTCDA height of lifting ($Z_{rel}$) from the Ag substrate, relative to the onset of the Kondo signature. The lifting step size was 20 pm.*

**The Raman maps at different vibrational modes**

The Raman spectroscopy was measured as a function of the lateral position of the tip above the PTCDA⁻ on two monolayers of NaCl. In addition to the map shown in Fig.3b, maps for other vibrational peaks and the background are presented in Fig.S8. The overall distributions of the intensity in the maps for individual modes are very similar, strongly indicating that the scattering process is resonant with the $D_0 \leftrightarrow D_3$ transition density. This interaction is strongest at the extremities of the molecule, reflecting that the scattering properties of the molecule are primarily driven by the coupling of excitonic transition with the localized plasmonic mode of the nanocavity (as explained above in the theoretical explanation of the map simulation).

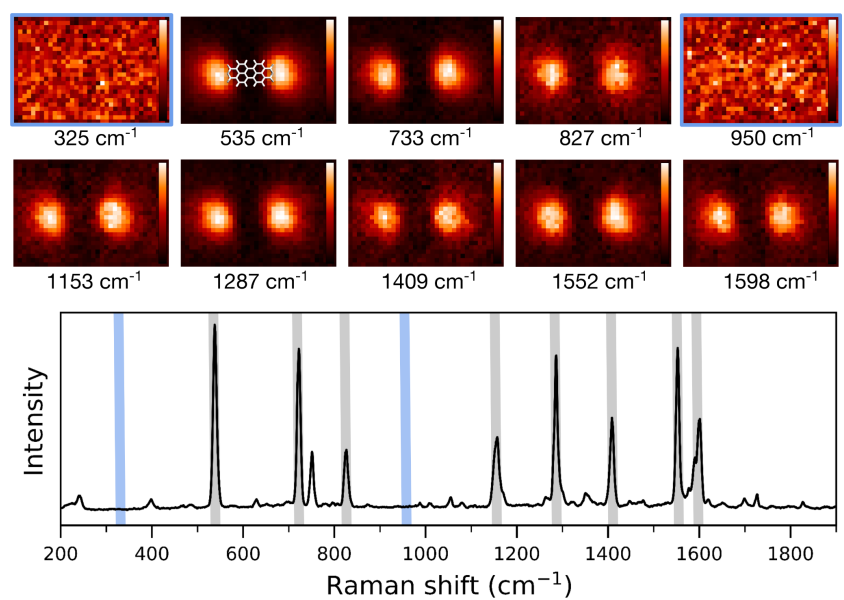

Fig.S8: *The constant-height TERS maps on the PTCDA on two layers of NaCl on Ag(111) (top), plotted for all dominant intensity peaks in the overall TERS spectrum (bottom). The size of all maps is 3.6 x 2.5 nm². Each map is normalized and the intensity is color-mapped using the same color-intensity scale in which the darkest tone corresponds to the minimum and the brightest to the maximum intensity value in a map. The spectra were measured at 0.5 V.*

**Supplementary Information References:**